\documentclass[nohyperref]{article}

\usepackage{microtype}
\usepackage{graphicx}
\usepackage{subfigure}
\usepackage{booktabs} % for professional tables

\usepackage{hyperref}
\usepackage{xurl}
\usepackage{xcolor}
\usepackage{amsmath}
\usepackage[T1]{fontenc}

\usepackage[accepted]{icml2022}

\usepackage{amsmath}
\usepackage{amssymb}
\usepackage{mathtools}
\usepackage{amsthm}

\usepackage[capitalize,noabbrev]{cleveref}

\usepackage[textsize=scriptsize,color=gray!20]{todonotes}

\begin{document}

\twocolumn[
% \icmltitle{The Technical and Scientific Structure of Emergent Language Research}
% \icmltitle{Modelling Methods of Emergent Language Research After Science and Engineering}
% \icmltitle{Beyond the Pioneering Phase of Emergent Language Research}
% \icmltitle{Building upon Exploratory Research in Emergent Language}
\icmltitle{Recommendations for Systematic Research on Emergent Language}




\begin{icmlauthorlist}
\icmlauthor{Brendon Boldt}{cmu}
\icmlauthor{David Mortensen}{cmu}
\end{icmlauthorlist}

\icmlaffiliation{cmu}{Language Technologies Institute, Carnegie Mellon University, Pittsburgh, PA 15213, USA}

\icmlcorrespondingauthor{Brendon Boldt}{bboldt@cs.cmu.edu}
\icmlcorrespondingauthor{David Mortensen}{dmortens@cs.cmu.edu}

\icmlkeywords{Machine Learning, ICML}

\vskip 0.3in
]



\printAffiliationsAndNotice{}  % leave blank if no need to mention equal contribution

\begin{abstract}
Emergent language is unique among fields within the discipline of machine learning for its open-endedness, not obviously presenting well-defined problems to be solved.
As a result, the current research in the field has largely been exploratory: focusing on establishing new problems, techniques, and phenomena.
Yet after these problems have been established, subsequent progress requires research which can measurably demonstrate how it improves on prior approaches.
This type of research is what we call \emph{systematic research}; in this paper, we illustrate this mode of research specifically for emergent language.
We first identify the overarching goals of emergent language research, categorizing them as either science or engineering.
Using this distinction, we present core methodological elements of science and engineering, analyze their role in current emergent language research, and recommend how to apply these elements.
\end{abstract}

\section{Introduction}

Emergent language research is characterized by the use of the neural networks and reinforcement learning to simulate the evolution of ``natural'' language from scratch (see \citet{lazaridou2020survey} for a comprehensive survey of emergent language research).
Both the range of potential applications of emergent language and the technique itself differ significantly from pre-existing methods in reinforcement learning and natural language processing.
Such novelty means that a large proportion of the work on the topic is exploratory research: focusing on establishing a baseline level of knowledge of new problems, techniques, and phenomena in light of the unique challenges of the field.

Yet after this foundation has been laid, subsequent long-term progress requires research which measurably demonstrates how it improves on prior approaches rather than tackling a new problem entirely.
This type of research is what we call \emph{systematic research}.
This systematicity is necessary for research to be effectively applied to the real problems and questions which motivate it in the first place.
The goal of this paper is to provide concrete recommendations for moving work on emergent language toward systematic research.

In order to develop these recommendations, we start in Section~\ref{sec:goals} by clarifying the goals of emergent language research; drawing attention to how they fit into the categories of science and engineering.
In Section~\ref{sec:elements-general} we illustrate the prototypical methods of science and engineering research which facilitate systematic research.
Finally, in Section~\ref{sec:elements-elr} we analyze to what extent current emergent language research uses these methods and recommend how they could be more fully employed.
The specific contributions of this work are:
\begin{enumerate}
    \item Recommending methods which move emergent language research toward systematic research, working towards overarching goals via measurable progress.
    \item Distinguishing between the capacities in which emergent language research can be science or engineering via the goals of a given project.
\end{enumerate}

\subsection{Exploratory and Systematic Research}

Exploratory research is research which focuses on innovative approaches and problems rather than pursuing a well-established goal.
Exploratory research can take place either in a new field, such as emergent language, or on the frontier of a more established research area.
The primary purpose of exploratory research is to build up a baseline level of knowledge of new approaches which differ significantly from well-studied approaches.
Once a critical mass of knowledge is acquired, a more structured and systematic research program is pursued in order to advance the knowledge and expertise of the field on that topic.

Systematic research is characterized by research contributions which measurably demonstrates how they improve on prior approaches; this entails adhering to established research problems.
The long-term success of any given research field is, in fact, predicated on systematic research; the direct comparison between two approaches to the same problem is what allows for significant progress to be made in an incremental way.
At times, ``incremental'' is considered a negative attribute, but this is far from the case in our own usage as the incremental nature of science and engineering are integral to their success as disciplines.
In the end, both exploratory and systematic research have a time and place where they are appropriate.

The ideas of exploratory and systematic research parallel the concepts of ``pre-paradigmatic science'' and ``normal science'' from \citet{kuhn_structures}.
Our account, though, is intended to apply to science \emph{and engineering} and is targeted towards practitioners of these fields rather philosophers of science.

\section{Goals of Emergent Language Research}%
\label{sec:goals}

The goals of a pursuit dictate the methods employed to achieve it.
Accordingly, we begin by enumerating the overarching goals of emergent language research.
Furthermore, these goals will also make the discussion of methods more concrete and easier to apply.

For our later discussion on methodology, it will prove fruitful to divide the goals of emergent language research into the categories of science and engineering.%
\footnote{The inspiration for this science-engineering distinction comes from the brief mention it was given in the abstract of \citet[p.~1]{lazaridou2020survey}.}
In the following subsections we enumerate first the scientific goals of emergent language research followed by the engineering goals.
This list is intended to be comprehensive breadth-wise although, on account of space, we are only able to offer a brief description of each goal.
For each goal, we present a handful of papers which have \emph{mentioned} the corresponding goal as a motivation for emergent language research.
Since most of the papers address relatively foundational topics, it is not as informative to look at what they directly address.

\subsection{Scientific Goals in Emergent Language}

Prototypically speaking, science is a descriptive pursuit seeking explanation and understanding of phenomena in the natural world \citep[p.~330]{Bunge_1966}.
Such explanations are to be found in scientific \emph{theories} built through empirical investigation \citep[abstract]{sep-structure-theory}.
The scientific goals of emergent language, then, center on its unique ability to support precise, controllable experiments on the formation, evolution, and use of language.
Since it is almost always impossible to run \textit{in situ} experiments with these phenomena, current investigation relies on observing historical data, making limited observations of real people, and small-scale laboratory studies.
The techniques of emergent language, then, are the sole hope to build theories of these aspects of human language with the typical methods of experimental science.

\paragraph{Evolution of Language}
Better understanding the origin and evolution of language is a straightforward line of scientific inquiry for emergent language because simulating these processes is, in fact, the distinguishing technique of this field.
The initial steps toward this goal are characterizing the rules governing the evolution of language abstractly from controllable emergent language experiments.
Subsequently the experiments can be made more realistic and adapted more directly to the evolution of human language.
Mentioned in
    \citet[p.~1]{lazaridou2020survey};
    \citet[p.~1]{lazaridou2018emergence};
    \citet[p.~1]{mordatch2018emergence};
    \citet[p.~1]{havrylov_emergence_2017}.

\paragraph{Building Blocks of Language}
The foundation of the structure and meaning of language include the linguistic disciplines of phonology, morphology, syntax, and semantics.
Emergent language can potentially serve as a testbed for disentangling the various aspects of language; something difficult to do via more traditional means like corpus-based or small-scale human studies.
Furthermore, if emergent language agents become fully competent language-users, it will an allow an unmatched degree of observability and controllability for the mechanics of language usage.
Mentioned in
    \citet[p.~1]{lazaridou2018emergence};
    \citet[p.~1]{bouchacourt_how_2018}.

\paragraph{Interactive Language}
Linguistic disciplines such as pragmatics and sociolinguistics focus on the external factors of language surrounding its use among multiple language users.
The interactive aspects of language are even more difficult to capture by static means (e.g., in corpora) than the aforementioned aspects, so the inherently interactive experiments within emergent language serve as an important tool for investigation here.
Language acquisition research, in particular, serves to greatly benefit from emergent language techniques due to the ability to both precisely observe pre-linguistic properties of language agents as well as easily test counterfactual hypotheses.
Mentioned in
    \citet[p.~1]{graesser2019emergent};
    \citet[p.~2]{eccles2019biases}.

\subsection{Engineering Goals in Emergent Language}
Engineering, in contrast to science, is a prescriptive pursuit as it recommends the best solutions to a problem rather than describing things as they are \citep[p.~330]{Bunge_1966}.
Specifically, engineering addresses problems which are well-defined insofar as it is possible to evaluate and compare competing solutions.
The engineering goals of emergent language research center around creating a language agent which is competent with an emergent language of comparable sophistication, complexity, or expressive power to human languages.
The end result is effectively having a computerized version of a human's language faculty which can be scaled, tweaked, and employed as needed for relevant machine learning applications.

\paragraph{Replication of Natural Language}
The core goal of emergent language research as engineering is to simulate the emergence of a language as complex and expressive as human language.
Most other scientific and engineering pursuits within emergent language research are predicated on making significant progress toward this goal.
The initial steps towards this goal subsume much of the current work which studies human language-like properties, like compositionality, in emergent languages.
Mentioned in \citet[p.~1]{evtimova_emergent_2018}; other papers discuss imitating aspects of human communication without explicitly mentioning the full replication of human language.

\paragraph{Alternative Data Paradigm}
The language agents developed through emergent language could serve as alternative paradigm for generating data for many other areas of natural language processing.
Since the language-competent agents are easily scalable and controllable (compared to humans), they could generate sophisticated synthetic data for the (pre)training and evaluation of machine learning models.
\citet{li2020translation} exemplify this by using synthetic data generated from an emergent language to improve the performance of a few-shot machine translation model.
Mentioned in
    \citet[p.~2]{lazaridou2016multiagent};
    \citet{li2020translation}.

Embodied NLP is an area of NLP that would reap an outsized benefit from emergent language as an alternative data source.
This is primarily because of the relative paucity of multi-modal---let alone interactive---data that the techniques of embodied NLP require.
The necessity of such multi-modal and interactive data for the ultimate goals of NLP are discussed in \citet[p.~8722]{bisk-etal-2020-experience}.

\paragraph{Communicating with Humans}
Simulating interactive and embodied language use facilitates improvements to the direct interaction between humans and artificial language agents.
Even outside of direct interaction, emergent language could augment the explainability of machine learning models.
Mentioned in
    \citet[p.~3]{lazaridou2020survey};
    \citet[p.~1]{chaabouni_antiefficient_2019};
    \citet[p.~1]{lowe_interaction_2020};
    \citet[p.~1]{lazaridou2016multiagent};
    \citet[p.~1]{havrylov_emergence_2017};
    \citet[p.~1]{kottur2017natural}.

\paragraph{Robust Multi-Agent Communication}
Apart from communicating directly with humans, there is a desire to imitate the generality and robustness which humans' use of language exhibits.
While emergent language is not the only path to this goal, its techniques potentially provide a more ``organic'' or ground-up way of developing such communication protocols rather than relying on hand-crafted methods of communication.
Mentioned in
    \citet[p.~3]{lazaridou2020survey};
    \citet[p.~1]{evtimova_emergent_2018};
    \citet[p.~1]{chaabouni2020compositionality};
    \citet[p.~2]{lazaridou2016multiagent};
    \citet[p.~1]{havrylov_emergence_2017};
    \citet[p.~1]{kottur2017natural}.

\paragraph{Improving Emergent Language Techniques}
Finally, an essential part of achieving all other goals of emergent language research is simply improving its own techniques.
This could come in a variety of forms including finding better ways to measure and analyze emergent languages, making certain properties more robust in emergent language experiments (e.g., compositionality), or comparing multiple approaches against each other.
Mentioned in
    \citet[p.~3]{lazaridou2020survey};
    \citet[p.~1]{chaabouni_antiefficient_2019};
    \citet[p.~1]{chaabouni2020compositionality};
    \citet[p.~1]{kharitonov_emergent_2020};
    \citet[p.~1]{cao_emergent_2018};
    \citet[p.~1]{eccles2019biases};
    \citet[p.~1]{resnick_capacity_2020};
    \citet[p.~1]{mordatch2018emergence}.

\subsection{Beyond Science and Engineering}
It is important to note that there are highly relevant \emph{philosophical} question surrounding emergent language, but since these fall outside the typical purview of the machine learning and computational linguistics communities, they are not incorporated into the analysis of this paper.
Such issues include providing a more rigorous and concrete definition for what experimental factors make an emergent language truly \emph{emergent} (as opposed to constrained or pre-planned) or how to quantify the expressivity or complexity of an emergent language vis-\`a-vis a human language.
As the scientific and engineering aspects of the field progress it will become clearer to what degree philosophically rigorous accounts are necessary, beyond pragmatic informal accounts determined by consensus.

\section{Methodological Elements}%
\label{sec:elements-general}

With the distinction between the scientific and engineering goals of emergent language in mind, we now address the \emph{methodological elements} of science and engineering.
These elements describe the concepts that are central to performing systematic research in the given discipline because they allow for research to measurably demonstrate how it improves over prior work.
They will serve as the structure of our analysis of current and future methods in emergent language research.
These elements are not intended to be necessary and/or sufficient conditions for science and engineering but rather prototypical elements which serve as a template.

\subsection{Elements of Science}

\emph{Theory} and \emph{theory-building} are the most pertinent elements of scientific research because they are both the means for achieving the explanatory goals as well as the measure of progress.
Thus, we present a collection of scientific elements which support theory-building, closely following definitions given in \citet{giere_science}.
This account of science seeks to avoid na\"ive, unphilosophical accounts as well as philosophically nuanced but unwieldy accounts of science.
Accordingly, we choose not to introduce science as a well-defined, five-step method because this is an oversimplification which fails to represent science as it is actually practiced \citep[Sec.~6.1]{sep-scientific-method}.

Two preliminary terms that we will use throughout this discussion are \emph{system} and \emph{object of study}.
We use system as a very general term denoting a collection of things which relate to each other in some reasonable way.
Systems range from natural systems like the solar system or an ecosystem to artificial systems like a car engine or an economy to conceptual systems like Euclidean geometry.
Secondly, when we talk about an object of study, we are referring to the particular system which a given line of inquiry seeks to understand, which may be different from the system which is the means of study.

\paragraph{Theoretical Model}
A \emph{theoretical model} (or \emph{theoretical framework}) is a conceptual system which is defined by a set of precise set of axioms or stipulations \citep[p.~80]{giere_science}.
% for the purpose explaining some natural system.
For example, Newton's three laws of motion define a system of moving and interacting physical bodies, and Mendel's theoretical model of inheritance defines a system of trait inheritance in sexually reproducing organisms.
These models have a tractable\footnote{``Tractable'' in the sense of being able to be understood by a human as opposed to some programmatically generated set of thousands or millions of axioms, suitable only for automated interpretation.} number of stipulations which can be used both for explanation as well as for predicting the behavior of the system which they describe.
Although theoretical models need not be formal in the strictest sense, there should be a precision about them that yields unambiguous conclusions given the axioms.

The core desideratum of a model is its accuracy: that its predictions are, in fact, borne out by experiments.
Nevertheless, there are a number of other characteristics which are important to models.
For example, the scope, simplicity, and even (controversially) beauty of a theoretical model can factor into how scientists evaluate it.

\paragraph{Computational Model}
\emph{Computational models} are similar to theoretical models insofar as both are systems \emph{defined} by a given set of stipulations.\footnote{As opposed something like a scale model which is not defined solely by a set of axioms}
Rather than providing a general-purpose description which can be used to reason and make predictions about a system, computational models define a procedure for simulating particular instances of some system.
Computational models, then, are often derived from theoretical models as a way of understanding the complex instances of the system which could not otherwise be reasoned about mentally or by hand.

\paragraph{Hypothesis and Experiment}
A \emph{hypothesis} is a statement about a system which can be either true or false \citep[p.~81]{giere_science}.
A hypothesis is always derived from a model of some sort.
This is straightforward in the case of a theoretical model which predicated on being able to develop predictions and hypotheses which can then be tested, but even when no explicit model is present, a hypothesis comes from some sort of implicit or intuitive model from which one makes an educated guess.

A good (not necessarily correct) hypothesis must be empirically testable, that is, one can formulate an \emph{experiment} which verifies or refutes the hypothesis by observing causes and effects in the actual system in question.
It is important to note that not all \emph{research questions} qualify as hypotheses; this does not make them inadequate \emph{as questions} but they cannot fulfill the role that hypotheses do in scientific research and theory-building in particular.

\paragraph{Theory-Building}
A \emph{theory} is a general hypothesis which states that the studied system is exactly or approximately described by a theoretical model \citep[p.~84]{giere_science}.
Since such a general hypothesis cannot be tested directly, we resort to using the theoretical model to generate smaller, more tractable hypotheses which can be directly tested.
These hypotheses are empirically tested; when a hypothesis is refuted, the model must be revised in order to account for this discrepancy with the system it claims to describe.
The modified model, then, is used to generate further hypotheses, forming an iterative process which is constitutive of \emph{theory-building}.
The measurable progress is scientific research is marked by theoretical models becoming more accurate, accounting for more phenomena, and better understood through this iterative process.

It is important to note that \emph{hypotheses which are not generated from theoretical models do not contribute to the process of building a theory}.
This is because when such a hypothesis is confirmed or rejected, that result does not, then, have anything to reflect back on.
Thus, since theory-building is critical to scientific research, it is necessary for hypotheses to be derived from theoretical models.

\subsection{Elements of Engineering}

Engineering research is directed towards solving well-defined problems.
Accordingly, one of the key elements of such research is the determining the quality of a solution to a problem.
The measurability of progress in engineering research is best illustrated by \emph{benchmarks} since they are the culmination of its other methodological elements.
The elements of engineering given here are selected and illustrated with computer vision serving as a prototype.

\paragraph{Problem Definition}
In engineering, it is integral to have a clear statement of the task to be accomplished as problem-solving is precisely the aim of the discipline; this we term the \emph{problem definition}.
A problem definition should prescribe the inputs and desired outputs of a solution and any constraints to be put on the solution (e.g., a fully self-driving car cannot require human input).

While a good problem definition is as precise as possible, many problems include aspects which are inherently difficult to define precisely.
For example, we could define the problem of image classification as,
    ``given an image $x$ depicting an instance of class $k$, the model should recover class $k$ solely from observing $x$.''
The definition of ``$x$ depicting $k$'' is inherently fuzzy and not formalizable.
Does a low-light, blurry image of a cheetah truly \emph{depict} a cheetah?
We argue that there is no precise answer to this question, yet the definition of image classification is still clear enough to serve as a productive problem definition.

\paragraph{Evaluation Metric}
A \emph{metric} is a quantitative measure of some property of a system.
An \emph{evaluation metric} is a metric which has a value judgment attached to it; that is, a metric which ought to be maximized (or minimized).
This valuation is rooted in the fact that it measures how well a proposed system solves the problem in question.
This is not to say that evaluation metrics always assign a better a score to a better solution to the problem, but insofar as the metric fails to do this, it is an imperfect evaluation metric.
Within computer vision, evaluation metrics can range from the straightforward, like accuracy for classification, to the highly engineered, like Fr\`echet inception distance for GAN image synthesis; in either case, the quality of a given solution to classification or synthesis is assessed through the use of these and other such evaluation metrics.
Without an evaluation metric, it is only possibly to qualitatively compare the goodness of competing solutions which is inherently less rigorous and repeatable than a quantitative assessment.

Metrics more generally do not necessarily have a value judgment attached to them.
For example, the slope of a linear regression is a metric but has no value in and of itself---it would not make sense to say the slope of one model is ``better'' than another without further context.

\paragraph{Standardization}
By \emph{standardization}, we mean the consistent use of a set of common problems, datasets, methods, evaluation metrics, etc.\@ within a field.
Within machine learning, standardization often occurs through consensus of the research community rather than via formal organizations.

Evaluating and comparing two different approaches to the same problem becomes difficult if not impossible unless most of the variables between the two are held constant.
In an extreme example, it would futile to attempt to compare two image classification architectures if one were trained on ImageNet and the other on CIFAR-10.
Nevertheless, even small differences in problem formulations can become confounding factors when comparing different approaches, necessitating some degree of standardization.

Although not integral to making comparisons between solutions, the standardization of tooling is also of great importance to engineering.
For example, within machine learning, the widespread use of neural network toolkits like TensorFlow or PyTorch make the development and evaluation of novel algorithms far easier as components and expertise are applicable across tasks and research groups.
Thus, standardized tooling pragmatically facilitates the rapid and wide diffusion of new techniques, greatly improving the quality and speed of engineering research.

\paragraph{Benchmark}
\emph{Benchmarks} are the result of tying together the standardization of a number of different aspects of a given engineering problem.
For machine learning, this typically includes the task formulation (e.g., model inputs and outputs), training and testing datasets, and evaluation metric.
Conversely benchmarks also \emph{encourage} standardization insofar as they provide pre-defined problem which is established as being relevant often with the attendant resources (e.g., dataset) needed to tackle it.
For example, the ImageNet \citep{imagenet} benchmark standardizes a dataset, problem formulation (image classification), and evaluation metric (top-$k$ accuracy) allowing for the direct comparison of image classifiers.

Due to the standardization of many facets of a given problem, benchmarks are a straightforward method of demonstrating how one approach measurably improves on a previous approach.
A measurable comparison may also be made without a formal benchmark, but the necessity of holding constant as many aspects of the problem as possible still applies.

\section{Elements of Emergent Language Research}%
\label{sec:elements-elr}
In this section, we will identify the role the aforementioned methodological elements play in emergent language research (according to its category) in the current body of research.
Building on this analysis of primarily exploratory research, we give recommendations on how to work towards more systematic research on emergent language.

\subsection{Science}

\subsubsection{Theoretical Model}
Theoretical models are employed only occasionally in current emergent language papers, although \citet{kottur2017natural,resnick_capacity_2020,kharitonov2020entropy} do employ theoretical models to some degree.
As an example we will look at the theoretical model presented by \citet{resnick_capacity_2020} which addresses relationship between model capacity and compositionality of emergent languages.

Specifically, an autoencoder with a discrete bottleneck layer is trained on a (finite) language $L$ generated by context-free grammar $G$.
The compositionality of the model is measured at the encoding yielded by the discrete bottleneck.
We could present a simplified\footnote{The original theoretical model does provide enough detail to apply directly to experiments.} version of the model as,
\begin{equation}
    \begin{split}
    |X| & \in [|G|, |L|) \wedge |Y| \not\in [|G|, |L|) \\
    & \Rightarrow C(X) > C(Y),
    \end{split}
\end{equation}
where
autoencoders $X$ and $Y$ have capacities $|X|$ and $|Y|$ (roughly parameter count),
    $C(\cdot)$ measures a trained model's bottleneck's compositionality,
    $|G|$ is the underfitting threshold,
    $|L|$ is the overfitting threshold,
    and $[|G|, |L|)$ is an interval.
If the autoencoder model's capacity is below $|G|$ (i.e., the sum of the terminal and non-terminal symbols and the production rules of $G$), the model will not be able to encode the language and underfit.
If the model's capacity is above $|L|$ (i.e., number of strings in the language), the model can simply memorize the strings (overfit) which will lead to a non-compositional protocol.
Thus, any model between the under- and over-fitting regimes achieves optimal performance by learning the underlying rules of the grammar, resulting in a higher compositionality than a model in one of those regimes.

The primary benefit to such a theoretical model is generating clear, testable hypotheses which is the basis for building a theory.
For example, the above model yields an easily testable hypothesis such as: for autoencoder $X$ with capacity $|X|=|G|+1$ and $Y$ with capacity $|Y|=|G|-1$, $C(X)>C(Y)$ (assuming $|G| + 1 < |L|$).

Looking forward in scientific emergent language research, the first step is to incorporate theoretical models more consistently in the research process.
Although not all claims can be neatly reduced to a handful of mathematical axioms, explicitly stating the model with as much specificity as possible is integral to generating informative hypotheses.
The second step is expanding the range of applicability or scope of theoretical models.
The model presented above is relatively limited with respect to the range of environments and agent architectures it can describe.
This reflects one of the greatest challenges for theoretical model-building in emergent language: the range of environments whence language can emerge is diverse, resisting simple formalization.
There are further ways in which to develop theoretical models, but addressing these two steps is sufficient for advancing beyond exploratory research on this front.

\subsubsection{Computational Model}%
\label{sec:elr-comp-model}
As a science, emergent language research typically has one of two systems as the object of study: either multi-agent communication in the abstract or human language as a part of nature.
It is important to be aware of this distinction due to the ubiquity of computational models in machine learning-based emergent language research.

When the studying multi-agent communication in the abstract, the computational model directly instantiates some theoretical system which is the object of study.
That is to say, the theoretical models and hypotheses concern the behavior of the computational model itself.
Using computational models in this way happens analogously in game theory, where we might search for Nash equilibria in a game by running simulations or search algorithms with a computer program.
This use of computational models encompasses a majority of current machine learning-based emergent language research.

If the ultimate object of study is human language itself, then a computational model is used as an approximation of real system of human language.
In this case, the theoretical model and hypotheses are \emph{actually} about the human use of language, and the computational model is only a proxy due to the inability to run experiments in the actual system\footnote{In contrast, in sciences like biology, chemistry, and physics, researchers can perform experiments directly in the system being studied (that is, nature).}.
In this mode of research, the computational models of emergent language have to imitate the particular circumstances of human language use as closely as possible in order for the computational model to serve as a good proxy, whereas this is not necessary if communication is only being studied in the abstract.

\subsubsection{Hypothesis and Experiment}
Hypotheses, in one form or another, show up about roughly half of the time in recent emergent language papers.
Exploratory research often contains research questions which are not hypotheses.
For example, \citet[p.~7]{evtimova_emergent_2018} phrase their research question as ``analyz[ing] the impact of changing the message dimensionality, and the effect of applying visual and linguistic attention mechanisms.''
Although this is an informative research question, it itself is not a prediction which can be shown to be true or false which is a necessary trait of a hypothesis factoring into the process of developing a theory.

When a hypothesis is present, it is typically derived from one of three sources:
    (1) from intuitions,
    (2) by analogy from a related field,
    or (3) from a theoretical model.
To say that a hypothesis comes from intuition is not to say it is random or entirely arbitrary, but it does mean that there is no set of explicit underlying assumptions that serve as the foundation for the hypothesis.
In other cases, a hypothesis is derived from an analogous field; for example, \citet[p.~1]{chaabouni2021color} study information content in a color naming game and derive their hypothesis (in part) from color-naming conventions observed in human languages.

Finally, hypotheses can be generated from a theoretical model, as illustrated in Section~\ref{sec:elr-comp-model}.
By combining the axioms, the model makes certain claims about how a given system will behave, which forms the core of a hypothesis.
A theoretical model facilitates deriving clear, testable hypotheses as the predictions and their preconditions are often evident looking at the theoretical model.
In most cases, if the aim is stated in the form of ``investigating the effect of $X$ on $Y$'' or ``seeing whether or not $X$'' occurs, it is not a testable hypothesis.
These aims are adequate for gaining a baseline sense of the stated topic, but since is not possible to confirm or refute such statements they cannot contribute to developing an underlying model.

\subsubsection{Theory}

Theories are not yet present in emergent language research primarily because there has only been sparse use of theoretical models, which form the core of a theory.
To move forward, then, it is necessary first to normalize the use of theoretical models in forming hypotheses and designing experiments.
After basic theoretical models are developed, it is then necessary to build on top of them by both amending them when their hypotheses are refuted and expanding their scope to account for a greater range of situations and phenomena.
Without taking an accumulative approach by building on top of prior theoretical models, it is difficult to determine if findings from the prior work applies to the subsequent approaches since any differences not accounted for by the model are confounding factors.

Expecting the construction of precise and broad-reaching theories in emergent language research analogous to what we see in the natural sciences like physics, biology, and chemistry is likely unreasonable.
First and foremost, this is because such a theory would entail a precise way to theoretically model reinforcement learning and neural networks, two fields dealing with complex systems undergoing rapid change and innovation.
The second difficulty is that when building theoretical models describing human language evolution, we do not even have access to the system we are trying to describe since we cannot directly observe or experiment with how human language evolves.
Both of these facts put emergent language research in a similar boat to scientific approaches to economics, first because economics fundamentally concerns the behavior of humans which is too complex to model precisely and second because it is impossible to run fully-fledged economic experiments (although economics is in a better position than emergent language research here).
Thus, it would be more reasonable to expect emergent language theories to develop more along the lines of scientific theories in economics, which do not provide natural science-level precision but still represent a rational and effective method of research.

\subsection{Engineering}

\subsubsection{Problem Definition}%
\label{sec:el-problem-definition}

Well-defined problems are occasionally given in the current body of emergent language literature and co-occur with topics that very clearly fall into the realm of engineering.
For example, \citet[p.~1]{bullard2021quasiequivalence} define a zero-shot communication class of problems which are defined by the following constraints in a cooperative game:
    (1) the agents train against themselves (self-play),
    (2) the agents are evaluated by playing with a new partner (cross-play),
    and (3) the symbols used to communicate have a variable cost assigned by the environment.
Further examples include \citet{eccles2019biases} and \citet{li2020translation} which offer formulations to particular, measurable problem.

To a certain extent, exploratory research is characterized by \emph{not} precisely defining problems because it is attempting to find that base level of knowledge which is necessary for a well-defined problem.
Thus, as more emergent language research moves to tackling real applications, it should become easier to precisely define problem a research project is addressing.
Nevertheless, being more intentional with defining and illustrating the problem addressed by the paper will catalyze this process.

\subsubsection{Evaluation Metric}
Similarly to problem definitions, evaluation metrics only appear occasionally in the current literature.
One group of metrics that has received considerable attention in the emergent language literature are metrics for measuring the compositionality of an emergent language.
In most cases, these are not true \emph{evaluation} metrics since they are not being optimized for as part of an engineering problem.
If such a metric were being optimized for, one could simply introduce strong inductive biases (e.g., hard-coding the language) towards a fully compositional protocol; although this may seem inappropriate, the constraints for the solution space must be clear from the problem definition.

Clear problem definitions go hand-in-hand with metrics which are truly evaluative as the criteria for success are clear in a well-defined problem.
For example, in the aforementioned papers in Section~\ref{sec:el-problem-definition}, we find that each problem has a clear evaluation metric appropriate for their problem definition:
    \citet{bullard2021quasiequivalence} study zero-shot coordination and evaluate with success rate,
    \citet{eccles2019biases} study the beneficial effect positive signalling between agents and evaluate with success rate and environment reward,
    and \citet{li2020translation} study emergent language for pretraining in few-shot translation and evaluate with BLEU.%

\subsubsection{Standardization}
Standard problems and approaches to problems are largely not present in emergent language research.
This is largely inherent to exploratory research since one of its primary purposes is to find approaches which are worthy of becoming standard approaches.

To a certain extent, consensus of a set of problem definitions will happen without needing explicit efforts.
Researchers will naturally choose the most salient problems to work on, and additionally, problem selection has a self-reinforcing effect where the more a given problem is researched, the more visibility it receives which in turn increases the likelihood of it being selected for subsequent studies.
On the other hand, novelty is often treated as inherently valuable in the publication process, which pushes against consensus and standardization.
Nevertheless, since other areas of machine learning have naturally standardized certain problems, it is reasonable to think that emergent language research can follow a similar path.

The best instance of standardization more generally in emergent language research is in the area of tooling: in particular, the EGG toolkit \citep{kharitonov-etal-2019-egg}.
This toolkit provides a simple way to write Python code for emergent language experiments. Such a resource can be especially helpful for emergent language researchers who do not have a strong background in computer programming.
It has been used in the implementation in a handful of papers (beyond the authors' own) including~\citet{sowik2020exploringsi,korbak2019developmentallyme,guo2020inductiveba}.

\subsubsection{Benchmark}

Benchmarks have not yet appeared in emergent language research.
This is due to the fact that benchmarks, in part, rely on a commonly agreed upon problem definition which is absent.
On the other hand, benchmarks can also be an impetus for standardization if the problem formulation and method of evaluation is of sufficient quality to convince a number of research groups to put their efforts towards this.
Nevertheless, it would be likely be difficult to build a convincing benchmark until some degree of consensus on worthwhile problem formulations is reached.

\subsection{Summary}

For the scientific goals of emergent language, the critical step towards systematic research is incorporating theoretical models into the standard research process.
Theoretical models will yield hypotheses and empirical evaluations which for the basis for building theories.
These theories are integral to understanding
    a number of questions about the nature language as well measurably gauging progress towards answering these questions.

For the engineering goals of emergent language, the critical step towards systematic research is precisely defining problems.
Well-defined problems form the basis for effective evaluation metrics and eventually standardization, the key factor in producing research contributions which measurably improve on prior approaches.

\section{Conclusion}

The above methodological recommendations offer a way to move work on emergent language toward systematic research, where research projects build off each other in a way that shows measurable improvement.
The exploratory research in the field thus far aids in recognizing the unique challenges of emergent language research which is necessary to implement these recommendations.
The distinction we have made between the capacities in which emergent language research can be science or engineering is critical---while the fields are closely related and intertwined, their paths to comparable, iterative research differ significantly.
Moving towards systematic research on emergent language will, then, facilitate the achievement of the unique and important goals of this field.

\subsubsection*{Acknowledgements}
This material is based on research sponsored in part by the Air Force Research Laboratory under agreement number FA8750-19-2-0200. The U.S.  Government is authorized to reproduce and distribute reprints for Governmental purposes notwithstanding any copyright notation thereon. The views and conclusions contained herein are those of the authors and should not be interpreted as necessarily representing the official policies or endorsements, either expressed or implied, of the Air Force Research Laboratory or the U.S. Government.

\bibliography{main}
\bibliographystyle{icml2022}

\end{document}